\documentclass[proceedings]{stacs}
\stacsheading{2010}{371-382}{Nancy, France}
\firstpageno{371}

\newcommand{\NP}{{\sf NP}}

\begin{document} 
\title[The $k$-in-a-path problem for claw-free graphs]{The $k$-in-a-path problem for claw-free graphs}

\author[lab1]{J. Fiala}{Ji\v{r}\'{\i} Fiala}
\address[lab1]{Charles University, Faculty of Mathematics and Physics,
        \newline DIMATIA and Institute for Theoretical Computer Science (ITI)
        \newline Malostransk\'e n\'am.~2/25, 118~00, Prague, Czech Republic}
\email{fiala@kam.mff.cuni.cz}
\email{bernard@kam.mff.cuni.cz}

\author[lab2]{M. Kami\'{n}ski}{Marcin Kami\'{n}ski}
\address[lab2]{Computer Science Department, Universit\'e Libre de Bruxelles,
	 \newline Boulevard du Triomphe CP212, B-1050 Brussels, Belgium}
\email{marcin.kaminski@ulb.ac.be}

\author[lab1]{B. Lidick\'{y}}{Bernard Lidick\'{y}}

\author[lab3]{D. Paulusma}{Dani\"el Paulusma}
\address[lab3]{Department of Computer Science, University of Durham,
        \newline Science Laboratories, South Road,
        \newline Durham DH1 3LE, England}
\email{daniel.paulusma@durham.ac.uk}      

\thanks{Research supported by the Ministry of Education of the Czech Republic as projects 1M0021620808 \newline and GACR~201/09/0197, by the Royal Society Joint Project Grant JP090172 and by EPSRC as EP/D053633/1.}
\keywords{induced path, claw-free graph, polynomial-time algorithm}
\subjclass{G.2.2 Graph algorithms, F.2.2 Computations on discrete structures}

\maketitle

\begin{abstract} 
Testing whether there is an induced path in a graph spanning $k$ given vertices is already \NP-complete in general graphs when $k=3$.
We show how to solve this problem in polynomial time on claw-free graphs, when $k$ is not part of the input but an arbitrarily fixed integer.
\end{abstract}

\section{Introduction}\label{s-intro}

Many interesting graph classes are closed under vertex deletion. Every such class can be characterized by a set of forbidden induced subgraphs.
One of the best-known examples is the class of perfect graphs.
A little over 40 years after Berge's conjecture, Chudnovsky et al.~\cite{CRST06} proved that  
a graph is perfect if and only if it contains neither an {\it odd hole} (induced cycle of odd length) nor an {\it odd antihole} (complement of an odd hole).
This motivates the research of detecting induced subgraphs such as paths and cycles, which is the topic of this paper. To be more precise,
we specify some vertices of a graph  called the {\it terminals} and
study the computational complexity of deciding if a graph has 
an induced subgraph of a certain type containing all the terminals. 
In particular, we focus on the following problem.

\medskip
\noindent
{\sc $k$-in-a-Path}\\
{\it Instance:} a graph $G$ with $k$ terminals.\\
{\it Question:} does there exist an induced path of $G$ containing 
the $k$ terminals?

\medskip
\noindent
Note that 
in the problem above, $k$ is a fixed integer. Clearly, the problem is polynomially solvable for $k=2$. 
Haas and Hoffmann~\cite{HH06} consider the case $k=3$.
After pointing out that this case is \NP-complete
as a consequence of a result by Fellows~\cite{Fe89}, they 
prove W$[1]$-completeness (where they take as parameter the 
length of an induced path that is a solution for {\sc $3$-in-a-Path}). 
Derhy and Picouleau~\cite{derhy2009} proved that the case $k=3$ 
is \NP-complete even for graphs with maximum degree at most three.

A natural question is what will happen if we relax the condition of ``being contained in an induced
path'' to ``being contained in an induced tree''.
This leads to the following problem.

\medskip
\noindent
{\sc $k$-in-a-Tree}\\
{\it Instance:} a graph $G$ with $k$ terminals.\\
{\it Question:} does there exist an induced tree of $G$ containing 
the $k$ terminals?

\medskip
\noindent
As we will see, also this problem has received a lot of attention in 
the last two years. It is \NP-complete if $k$ is part of the input~\cite{derhy2009}. However, 
Chudnovsky and Seymour~\cite{CS} have recently given a deep and complicated polynomial-time algorithm for  the case $k=3$.

\begin{theorem}[\cite{CS}]\label{t-cs}
The {\sc $3$-in-a-Tree} problem is solvable in polynomial time.
\end{theorem}

The computational complexity of {\sc $k$-in-a-Tree} for $k=4$ is still open. So far, only partial results are known, 
such as a polynomial-time algorithm for $k=4$ when the input is 
triangle-free by Derhy, Picouleau and Trotignon~\cite{DPT09}. 
This result and Theorem~\ref{t-cs} 
were extended by Trotignon and Wei~\cite{TW09} who showed that 
{\sc $k$-in-a-Tree} is polynomially solvable 
for graphs of girth at least $k$. The authors of~\cite{DPT09} also show that it
is \NP-complete to decide if a graph $G$ contains an induced tree $T$ covering four specified vertices such that $T$ has at most one vertex of degree at least three. 

In general, {\sc $k$-in-a-Path} and {\sc $k$-in-a-Tree} are only equivalent 
for $k\leq 2$. However, in this paper, we study 
{\it claw-free} graphs (graphs
with no induced 4-vertex star).
Claw-free graphs are a rich and well-studied class containing, e.g., the class
of (quasi)-line graphs and the class of complements of triangle-free graphs;
see~\cite{FFR97} for a survey. Notice that any induced tree in a claw-free graph is in fact an induced path.

\begin{observation}~\label{o-ob1}
The {\sc $k$-in-a-Path} and {\sc $k$-in-a-Tree} problem are 
equivalent for the class of claw-free graphs.
\end{observation}

\noindent
{\bf Motivation.}
The polynomial-time algorithm for {\sc 3-in-a-Tree}~\cite{CS} 
has already proven to be a powerful tool for several problems. For instance, 
it is used as a subroutine in polynomial time algorithms for detecting
induced thetas and pyramids~\cite{CS} and several 
other induced subgraphs~\cite{LMT07}.
The authors of~\cite{HKP09} use it  to solve the {\sc Parity Path} problem in polynomial time for claw-free graphs.
(This problem is to test if a graph contains both an odd and even length induced paths between two specified vertices.
It is \NP-complete in general as shown by Bienstock~\cite{Bi91}.)

L\'{e}v\^{e}que et al.~\cite{LMT07} use the algorithm of~\cite{CS} to solve
the $2$-{\sc Induced Cycle} problem in polynomial time for graphs not containing an induced path or 
subdivided claw on some fixed number of vertices. 
The $k$-{\sc Induced Cycle} problem is to test if a graph contains an induced cycle spanning $k$
terminals. In general it is \NP-complete already for $k=2$~\cite{Bi91}.
For fixed $k$,
an instance of this problem can be reduced
to a polynomial number of
instances of the {\sc $k$-Induced Disjoint Paths} problem, which we define below.
Paths $P_1,\ldots, P_k$ in a graph $G$ 
are said to be {\it mutually induced} if 
for any $1\leq i<j \leq k$, $P_i$ and $P_j$ 
have neither common vertices 
(i.e. $V(P_i)\cap V(P_j)=\emptyset$) 
nor adjacent vertices (i.e. $uv\notin E$ for any $u\in V(P_i), v\in V(P_j)$).

\medskip
\noindent
{\sc $k$-Induced Disjoint Paths}\\
{\it Instance:} a graph $G$ with $k$ pairs of terminals $(s_i,t_i)$ for
$i=1,\ldots,k$.\\
{\it Question:} does $G$ contain $k$ mutually induced paths $P_i$ such
that $P_i$ connects $s_i$ and $t_i$ for $i=1,\ldots,k$?

\medskip
\noindent
This problem is \NP-complete for $k=2$~\cite{Bi91}. Kawarabayashi and Kobayashi~\cite{KK08} showed that, for any fixed $k$, the {\sc $k$-Induced Disjoint Paths} problem is solvable in linear time on planar graphs and that consequently
{\sc $k$-Induced Disjoint Cycle} is solvable in polynomial time on this graph 
class for any fixed $k$. In~\cite{KK09},
Kawarabayashi and Kobayashi improve the latter result by presenting 
a linear time algorithm for this problem,
and even extend the results for both these problems to graphs of bounded genus. 
As we shall see, we can also solve {\sc $k$-Induced Disjoint Paths} and $k$-{\sc Induced Cycle} in polynomial time in claw-free graphs.
The version of the problem in which any two paths are vertex-disjoint but
may have adjacent vertices is called the {\sc $k$-Disjoint Paths} problem.
For this problem Robertson and Seymour~\cite{RS95} proved the following result.

\begin{theorem}[\cite{RS95}]\label{t-RS}
For fixed $k$, the {\sc $k$-Disjoint Paths} problem is solvable in polynomial time.
\end{theorem}

\noindent
{\bf Our Results and Paper Organization.}
In Section~\ref{s-pre} we define some basic terminology. Section~\ref{s-main} contains our main result:  {\sc $k$-in-a-Path} is solvable in polynomial time in claw-free graphs
for any fixed integer $k$. This, in fact, follows from a stronger theorem proved in Section~\ref{s-proof}; the problem is solvable in polynomial time even if the terminals are to appear on the path in a fixed order. A consequence of our result is that the {\sc $k$-Induced Disjoint Paths} and {\sc $k$-Induced Cycle} problems are polynomially solvable in claw-free graphs 
for any fixed integer $k$. 
In Section~\ref{s-proof} we present our polynomial-time algorithm that solves the ordered version of {\sc $k$-in-a-Path}. The algorithm first performs ``cleaning of the graph''. This is an operation introduced in~\cite{HKP09}.  After cleaning the graph is free of odd antiholes of length at least seven. Next we treat odd holes of length five that are contained in the neighborhood of a vertex. The resulting graph is quasi-line. Finally, we solve the problem using a recent characterization of quasi-line graphs by Chudnovsky and Seymour~\cite{CS05} and related algorithmic results of King and Reed~\cite{KR08}. In Section~\ref{s-con} we mention relevant open problems.

\section{Preliminaries}\label{s-pre}

All graphs in this paper are undirected, finite, and neither have loops nor multiple edges.
Let $G$ be a graph. 
We refer to the vertex set and edge set of $G$ by $V=V(G)$ and $E=E(G)$, respectively. 
The {\em neighborhood} of a vertex $u$ in $G$ is denoted by $N_G(u)=\{v\in V\ |\ uv\in E\}$.
The subgraph of $G$ induced by $U\subseteq V$ is denoted $G[U]$.
Analogously, the {\it neighborhood} of a set $U\subseteq V$ is $N(U):=\bigcup_{u\in U} N(u) \setminus U$.
We say that two vertex-disjoint subsets of $V$ are {\it adjacent} if some of their vertices are adjacent. 
The {\it distance} $d(u,v)$ between two vertices $u$ and $v$
in $G$ is the number of edges on a shortest
path between them.
The {\it edge contraction} of an edge $e=uv$ removes its two end vertices $u,v$ and replaces it by a new
vertex adjacent to all vertices in $N(u)\cup N(v)$ (without introducing loops or multiple edges).

We denote the path and cycle on $n$ vertices by $P_n$ and $C_n$, respectively.
Let $P=v_1v_2\ldots v_p$ be a
path with a fixed orientation. The successor $v_{i+1}$ of $v_i$ is
denoted by $v_i^+$ and its predecessor $v_{i-1}$ by
$v_i^-$. The segment $v_iv_{i+1} \ldots v_j$ is denoted by
$v_i\overrightarrow{P}v_j$. The converse segment $v_jv_{j-1} \ldots v_i$ is
denoted by $v_j\overleftarrow{P}v_i$.
 
A {\em hole} is an induced cycle of length at least 4 and an {\em antihole} is the complement of a hole. 
We say that a hole is {\em odd}  
if it has an odd number of edges. 
An antihole is called odd 
if it is the complement is an odd hole.

A {\em claw} is the graph $(\{x,a,b,c\},\{xa,xb,xc\})$, where vertex $x$ is called the {\em center} of the claw.
A graph is {\it claw-free} if it does not contain a claw as an induced subgraph.
A \emph{clique} is a subgraph isomorphic to a complete graph.
A {\em diamond} is a graph obtain from a clique on four vertices after removing one edge.
A vertex $u$ in a graph $G$ is {\it simplicial} if $G[N(u)]$ is a clique.

Let $s$ and $t$ be two specified vertices in a graph $G=(V,E)$.
A vertex $v\in V$ is called {\it irrelevant} for vertices $s$ and $t$ if 
$v$ does not lie on any induced path from $s$ to $t$. A graph $G$ is {\it clean} if none of its vertices is irrelevant.
We say that we {\it clean} $G$ for $s$ and $t$ by repeatedly deleting irrelevant vertices for $s$ and $t$ as long as possible.
In general, determining if a vertex is irrelevant is NP-complete~\cite{Bi91}. 
However, for claw-free graphs, the authors of~\cite{HKP09} could show the following (where they used
Observation~\ref{o-ob1} and Theorem~\ref{t-CS} for obtaining the polynomial time bound).

\begin{lemma}[\cite{HKP09}]\label{l-clean}
Let $s,t$ be two vertices of a claw-free graph $G$.
Then $G$ can be cleaned for $s$ and $t$ in polynomial time.
Moreover, the resulting graph does not contain an odd antihole of length at least seven.
\end{lemma}

The {\it line graph\/} of a graph $G$ with edges $e_1,\ldots,e_p$ is
the graph $L=L(G)$ with vertices $u_1,\ldots,u_p$ such that there is
an edge between any two vertices $u_i$ and $u_j$ if and only if
$e_i$ and $e_j$ share an end vertex in $H$.
We note that mutually induced paths in a line graph $L(G)$ 
are in one-to-one correspondence with vertex-disjoint paths in $G$.
Combining this observation with Theorem~\ref{t-RS} leads to the following
result.

\begin{corollary}\label{c-easy}
For fixed $k$, the {\sc $k$-Induced Disjoint Paths} problem can be solved in
polynomial time in line graphs.
\end{corollary}

A graph $G=(V,E)$ is called a \emph{quasi-line graph} if
for every vertex $u\in V$ there exist two vertex-disjoint cliques $A$ and $B$ in $G$
such that $N(u)=V(A)\cup V(B)$ (where $V(A)$ and $V(B)$ might be adjacent).
Clearly, every line graph is quasi-line and every quasi-line graph is claw-free.
The following observation is useful and easy to see by looking at the complements of neighborhood in a graph. 

\begin{observation}\label{o-odd}
A claw-free graph $G$ is a quasi-line graph if and only if $G$ does not 
contain a vertex with an odd antihole in its neighborhood.
\end{observation}

A clique in a graph $G$ is called 
{\it nontrivial} if it contains at least two vertices. A nontrivial
clique $A$ is called {\it homogeneous} if 
every vertex in $V(G)\backslash V(A)$ is either adjacent to all vertices of $A$ or to none of them. Notice that it is possible to check in polynomial time if an edge of the graph is a homogeneous clique.  This justifies the following observation.

\begin{observation}\label{o-h}
The problem of detecting a homogeneous clique in a graph is solvable in polynomial time.
\end{observation}

Two disjoint cliques $A$ and $B$ form a \emph{homogeneous pair} in $G$ if 
the following two conditions hold. First, at least one of $A,B$ contains more than one vertex. Second,
every vertex $v\in V(G)\setminus(V(A) \cup V(B))$ is
either adjacent to all vertices of $A$ or to none vertex of $A$
as well as either adjacent to all of $B$ or to none of $B$.
The following result by King and Reed~\cite[Section 3]{KR08} will be useful.

\begin{lemma}[\cite{KR08}]\label{l-king}
The problem of detecting a  homogeneous pair of cliques in a 
graph is solvable in polynomial time.
\end{lemma}

Let $V$ be a finite set of points of a real line, and ${\mathcal I}$ be a collection of intervals.
Two points are adjacent if and only if they belong to a common interval $I\in {\mathcal I}$.
The resulting graph is a \emph{linear interval graph}. 
Analogously, if we consider a set of points of a circle and set of intervals (angles) on the circle
we get a \emph{circular interval graph}.
Graphs in both classes are claw-free, in fact linear interval graphs coincide with 
proper interval graphs (intersection graph of a set of intervals on a line, where no 
interval contains another from the set) and circular interval graphs coincide with
proper circular arc graphs (defined analogously). We need the following result of Deng, Hell, and Huang~\cite{DHH96}.

\begin{figure}
 \centering
   \includegraphics[scale=.75]{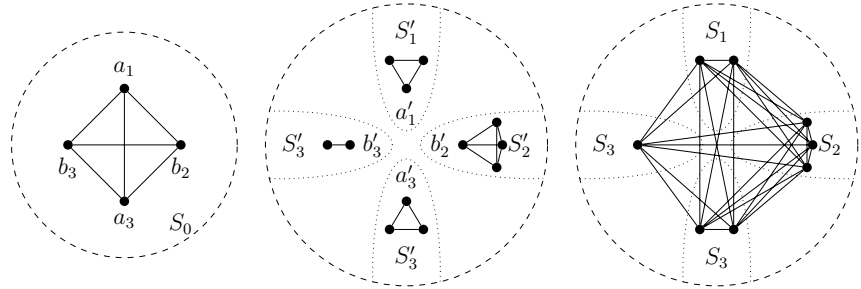}
  \caption{Composition of three linear interval strips (only part of the graph is displayed).}\label{f-gprime}
\end{figure}

\begin{theorem}[\cite{DHH96}]\label{t-deng}
Circular interval graphs and linear interval graphs can be recognized 
in linear time. Furthermore, a corresponding representation of such graphs can be constructed in linear time
as well.
\end{theorem}

A \emph{linear interval strip} $(S,a,b)$ is a linear interval graph $S$ where 
$a$ and $b$ are 
the leftmost and the rightmost points (vertices) of its representation.
Observe that in such a graph the vertices $a$ and $b$ are simplicial.
Let $S_0$ be a graph with vertices $a_1,b_1,\dots,a_n,b_n$ that is isomorphic 
to an arbitrary disjoint union of complete graphs.
Let $(S_1',a_1',b_1'),\dots,(S_n',a_n',b_n')$ be a collection of linear interval strips.
The \emph{composition} $S_n$  is defined inductively where $S_i$ is formed from the disjoint
union of $S_{i-1}$ and $S_i'$, where:
\begin{itemize}
\item [$\bullet$] all neighbors of $a_i$ are connected to all neighbors of $a_i'$;
\item [$\bullet$] all neighbors of $b_i$ are connected to all neighbors
of $b_i'$;  
\item [$\bullet$] vertices $a_i,a_i',b_i,b_i'$ are removed.
\end{itemize}
See Figure~\ref{f-gprime} for an example.
We are now ready to state the structure of quasi-line graphs as characterized by Chudnovsky and Seymour~\cite{CS05}.

\begin{theorem}[\cite{CS05}]\label{t-CS}
A quasi-line graph $G$ with no homogeneous pair of cliques 
is either a circular interval graph or a composition of linear interval strips. 
\end{theorem}

Finally, we need another algorithmic result of King and Reed~\cite{KR08}.
They observe that 
the composition of the final strip in a composition of linear interval graphs
is a so-called nontrivial interval 2-join and that every nontrivial interval
2-join contains a so-called canonical interval 2-join. 
In Lemma 13 of this paper 
they show how to find in polynomial time a canonical interval 2-join in a quasi-line graph with no homogeneous pair of cliques and
no simplicial vertex or else to conclude that none exists. Recursively applying
this result leads to the following lemma.

\begin{lemma}[\cite{KR08}]\label{l-king2}
Let $G$ be a quasi-line graph with no homogeneous pairs of cliques and
no simplicial vertex that is a composition of linear interval strips. 
Then the collection of linear
interval strips that define $G$ can be found in polynomial time. \end{lemma}

\section{Our Main Result}\label{s-main}

Here is our main result.

\begin{theorem}\label{t-main}
For any fixed  $k$, the {\sc $k$-in-a-Path} problem is solvable in 
polynomial time in claw-free graphs.
\end{theorem}

In order to prove Theorem~\ref{t-main} we define the following problem.

\medskip
\noindent
{\sc Ordered-$k$-in-a-Path}\\
{\it Instance:} a graph $G$ with $k$ terminals ordered as $t_1,\ldots,t_k$.\\
{\it Question:} does there exist an induced path of $G$ starting in $t_1$ then passing through
$t_2,\ldots,t_{k-1}$ and ending in $t_k$?

\medskip
\noindent
We can resolve the original {\sc $k$-in-a-Path} problem by $k!$ rounds of the more specific version defined above, 
where in each round we order the terminals by a different permutation. Hence, since we assume that $k$
is fixed, it suffices to prove Theorem~\ref{t-korder} in order to 
obtain Theorem~\ref{t-main}.

\begin{theorem}\label{t-korder}
For any fixed $k$, the {\sc Ordered-$k$-in-a-Paths} problem is solvable in polynomial time in claw-free graphs.
\end{theorem}

We prove Theorem~\ref{t-korder} in Section~\ref{s-proof} and finish
this section with the following consequence of it.

\begin{corollary}\label{c-pacy}
For any fixed $k$, the {\sc $k$-Disjoint Induced Paths} and
{\sc $k$-Induced Cycle} problem are solvable in polynomial time in 
claw-free graphs.
\end{corollary}

\begin{proof}
Let $G$ be a claw-free graph that together with terminals $t_1,\ldots,t_k$ 
is an instance of {\sc $k$-Induced Cycle}. We fix an order of the terminals,
say, the order is $t_1,\ldots,t_k$. We fix 
neighbors $a_i$ and $b_{i-1}$ of each terminal $t_i$.
This way we obtain an instance of {\sc $k$-Induced Disjoint Paths} with pairs of terminals
$(a_i,b_i)$ where $b_0=b_k$. 
Clearly, the total number of instances we have created is polynomial.
Hence, we can solve {\sc $k$-Induced Cycle} in polynomial time
if we can solve {\sc $k$-Induced Disjoint Paths} in polynomial time.

Let $G$ be a claw-free graph that together with $k$ pairs of terminals 
$(a_i,b_i)$ for $i=1,\ldots,k$ is an instance of the {\sc $k$-Induced Disjoint Paths} problem.  
First we add an edge between each pair of non-adjacent neighbors of every
terminal in $T=\{a_1,\ldots, a_k, b_1,\ldots, b_k\}$. We denote the resulting
graphs obtained after performing this operation on a terminal by 
$G_1,\ldots,G_{2k}$, and define $G_0:=G$. 
We claim that $G'=G_{2k}$ is claw-free and prove this by induction.

The claim is true for $G_0$. Suppose the claim is true for $G_j$ for some
$0\leq j\leq 2k-1$. Consider $G_{j+1}$ and suppose, for contradiction, that 
$G_{j+1}$ contains an induced subgraph isomorphic to a claw. Let 
$K:=\{x,a,b,c\}$ be a set of vertices of $G_{j+1}$ inducing a claw with center $x$. Let $s\in T$ be the vertex of $G_j$ that becomes simplicial in $G_{j+1}$.
Then $x\neq s$. Since 
$G_j$ is claw-free, we may without loss of generality assume that at least two vertices of $K$ must be in $N_{G_{j+1}}(s)\cup \{s\}$. Since $N_{G_{j+1}}(s)\cup \{s\}$ is a clique of $G_{j+1}$ and $\{a,b,c\}$ is an independent set of 
$G_{j+1}$, we may without loss of generality assume that $K\cap (N_{G_{j+1}}(s)\cup \{s\})=\{x,a\}$ and $\{b,c\}\subseteq V(G_{j+1})\setminus (N_{G_{j+1}}(s)\cup \{s\})$. Then $\{x,b,c,s\}$ induces a claw in $G_j$ with center $x$, a contradiction. Hence, $G'$ is indeed claw-free.

We note that $G$ with terminals $(a_1,b_1),\ldots,(a_k,b_k)$ forms a {\sc Yes}-instance of {\sc $k$-Induced Disjoint Paths} if and only if $G'$ with
the same terminal pairs is a {\sc Yes}-instance of this problem.
In the next step we identify terminal $b_i$ with $a_{i+1}$, i.e., for 
$i=1,\ldots, k-1$ we remove $b_i, a_{i+1}$ and replace them by a new vertex $t_{i+1}$ adjacent to all neighbors of 
$a_{i+1}$ and to all neighbors of $b_i$. We call the resulting graph $G''$ and
observe that $G$ is claw-free.
We define $t_1:=a_1$ and $t_{k+1}:=b_k$ and claim that 
$G'$ with terminal pairs $(a_1,b_1),\ldots, (a_k,b_k)$ forms a {\sc Yes}-instance of the {\sc $k$-Induced Paths} problem if and only
if $G''$ with terminals $t_1,\ldots, t_{k+1}$ forms a {\sc Yes}-instance of the 
{\sc Ordered-$(k+1)$-in-a-Path} problem. 

In order to see this, suppose $G'$ contains $k$ mutually induced paths $P_i$ such
that $P_i$ connects $a_i$ to $b_i$ for $1\leq i \leq k$. 
Then 
$$P=
t_1\overrightarrow{P_1}b_1^-t_2a_2^+\overrightarrow{P_2}b_2^-\ldots 
t_ka_k^+\overrightarrow{P_k}t_k$$ is an induced path passing through
the terminals $t_i$ in prescribed order.
Now suppose $G''$ contains an induced path $P$ passing through terminals
in order $t_1,\ldots,t_{k+1}$. For $i=1,\ldots,k+1$ we define 
paths
$P_i=a_it_i^+\overrightarrow{P}t_{i+1}^-b_i$, which 
are mutually induced. We now 
apply Theorem~\ref{t-korder}. 
This completes the proof. 
\end{proof}

\section{The Proof of Theorem~\ref{t-korder}}\label{s-proof}

We present a polynomial-time algorithm that solves the 
{\sc Ordered-$k$-in-a-Path} problem on  a claw-free graph $G$ with terminals
in order $t_1,\ldots, t_k$ for any fixed integer $k$. 
We call an induced path $P$ from $t_1$ to $t_k$ that contains the other terminals in order $t_2,\ldots,t_{k-1}$ 
a {\it solution} of this problem.
Furthermore, an operation in this algorithm on input graph $G$ with terminals $t_1,\ldots,t_k$ 
{\it preserves the solution} if the following holds: the resulting graph $G'$ with resulting terminals
$t_1',\ldots, t_{k'}'$ for some $k'\leq k$ is a {\sc Yes}-instance
of the {\sc Ordered-$k'$-in-a-Path} problem
if and only if $G$ is a {\sc Yes}-instance of the {\sc Ordered-$k$-in-a-Path} problem. 
We call $G$ {\it simple} if the following three conditions hold:

\begin{itemize}
\item [(i)] $t_1,t_k$ are of degree one in $G$ and all other terminals 
$t_i$ ($1 < i < k$) are of degree two in $G$, and the two 
neighbors of such $t_i$ are not adjacent;
\item [(ii)] the distance between any pair $t_i,t_j$ is at least four;
\item [(iii)] $G$ is connected.
\end{itemize}

\noindent \rule{\textwidth}{0.1mm}\\
{\sc The Algorithm and Proof of Theorem~\ref{t-korder}}

\medskip
\noindent
Let $G$ be an input graph with terminals $t_1,\ldots,t_k$. 

\medskip
\noindent
If $k=2$, we compute a shortest path from $t_1$ to $t_2$. If $k=3$, we use Theorem~\ref{t-cs} together with Observation~\ref{o-ob1}. Suppose $k\geq 4$.

\medskip
\noindent
{\bf Step 1. Reduce to a set of simple graphs.}\\
We apply Lemma~\ref{l-simple} and obtain in polynomial time a set ${\mathcal G}$
that consists of a polynomial number
of simple graphs of size at most $|V(G)|$ such that there is a solution for $G$
if and only if there is a solution for one of the graphs in ${\mathcal G}$.
We consider {\it each} graph in ${\mathcal G}$. 
For convenience we denote such a graph by $G$ as well.

\medskip
\noindent
{\bf Step 2. Reduce to a quasi-line graph.}\\
We first clean $G$ for $t_1$ and $t_k$. If during cleaning we remove a terminal, then we output {\sc No}. Otherwise, clearly,
we preserve the solution. By Lemma~\ref{l-clean}, this can be done in polynomial time and ensures that there are no
odd antiholes of length at least seven left. Also, $G$ stays simple.
Then we apply Lemma~\ref{l-c5}, which removes vertices $v$ whose neighborhood contain an
odd hole of length five, as long as we can. Clearly, we can do this in  polynomial time. 
Note that $G$ stays connected since we do not remove cut-vertices due to the claw-freeness.
By condition (i), we do not remove a terminal either.
Afterwards, we clean $G$ again for $t_1$ and $t_k$. 
If we remove a terminal, we output {\sc No}. Otherwise, as a result of our operations, $G$ becomes a simple quasi-line graph
due to Observation~\ref{o-odd}.

\medskip
\noindent
{\bf Step 3. Reduce to a simple quasi-line graph with no homogeneous clique}\\
We first exhaustively search for homogeneous cliques by running the
polynomial algorithm mentioned in Observation~\ref{o-h}
and apply Lemma~\ref{l-hc} each time we find such a clique. Clearly, we can
perform the latter in polynomial time as well.
After every reduction of such a clique to a single vertex, 
$G$ stays simple and quasi-line, and at some moment does not contain any homogeneous clique anymore, while we preserve the solution.

\medskip
\noindent
{\bf Step 4. Reduce to a circular interval graph or to a composition of interval strips.}\\
Let $t_1',t_k'$ be the (unique) neighbor of $t_1$ and $t_k'$, respectively.
As long as $G$ contains homogeneous pairs of cliques $(A,B)$ so that
$A$ neither $B$ is equal to $\{t_1,t_1'\}$ or $\{t_k,t_k'\}$, we do as follows.
We first detect such a pair in polynomial time using Lemma~\ref{l-king} 
and reduce them to a pair of single vertices by applying Lemma~\ref{l-noterminal}.  Also performing Lemma~\ref{l-noterminal} clearly takes only polynomial time. After every reduction, $G$ stays simple and quasi-line, and
we preserve the solution. At some moment, the only homogeneous pairs of cliques that are possibly left in $G$ are of
the form $(\{t_1,t_1'\},B)$ and $(\{t_k,t_k'\},B)$. As $G$ does not contain 
a homogeneous clique (see Step 3), 
the cliques in such pairs must have adjacent vertex sets.
Hence, there can be at most two of such pairs. We perform 
Lemma~\ref{l-noterminal} and afterwards make the graph simple again.
Although this might result in a number of new instances, their total number is still polynomial because we perform this operation at most twice. Hence, we may
without loss of generality assume that $G$ stays simple.
By Theorem~\ref{t-CS}, $G$ is either a circular interval graph or a composition of linear interval strips; we deal with theses two cases separately after
recognizing in polynomial time in which case we are 
by using Theorem~\ref{t-deng}.

\medskip
\noindent
{\bf Step 5a. Solve the problem for a circular interval graph.}\\
Let $G$ be a circular interval graph.
Observe that the order of vertices in an induced path must respect the natural order of points on a circle.
Hence, deleting all points that lie on the circle between $t_k$ and $t_1$
preserves the solution. So, we may even assume that $G$ is a linear interval graph. We solve the problem in these graphs in Theorem~\ref{t-lin}.

\medskip
\noindent
{\bf Step 5b. Solve the problem for a composition of linear interval strips.}\\
Let $G$ be a composition of linear interval strips. Because $G$ is assumed
to be clean for $t_1,\ldots,t_k$, $G$ contains no simplicial vertex.
Then we can find these
strips in polynomial time using Lemma~\ref{l-king2} and use this information in
Lemma~\ref{l-tol}. There we create a line graph $G'$ with $|V(G')|\leq 
|V(G)|$, while preserving the solution.
Moreover, this can be done in polynomial time by the same theorem. 
Then we use Corollary~\ref{c-easy} to prove that the problem is 
polynomially solvable in line graphs in Theorem~\ref{t-linegraph}.

\smallskip
\noindent \rule{\textwidth}{0.1mm}

\medskip
Now it remains to state and prove Lemmas~\ref{l-simple}--\ref{l-tol}
and Theorems~\ref{t-lin}--~\ref{t-linegraph}.

\begin{lemma}\label{l-simple}
Let $G$ be a graph with terminals ordered $t_1,\ldots, t_k$. 
Then there exists a set ${\mathcal G}$ of $n^{O(k)}$ simple graphs, each of size
at most $|V(G)|$, such that $G$ has a solution if and only if 
there exists a graph in ${\mathcal G}$ that has a solution.
Moreover, ${\mathcal G}$ can be constructed in polynomial time.
\end{lemma}

\begin{proof}
We branch as follows.
First we guess the first six vertices after $t_1$ in a possible solution.
Then we guess the last six vertices before $t_n$. Finally, for 
$2\leq i\leq n-1$, we guess the last six vertices preceding $t_i$ and the first
six vertices following $t_i$. We check if the subgraph induced by the terminals and  all guessed vertices has maximum degree 2. If not we discard this guess.
Otherwise, for every terminal and for every guessed vertex
that is not an end vertex of a guessed subpath, 
we remove all its neighbors that are not guessed
vertices. This way we obtain a number of graphs 
which we further process one by one.

Let $G'$ be such a created subgraph.
If $G'$ does not contain all terminals, we discard $G'$. 
If $G'$ is disconnected then 
we discard $G'$ if two terminals are in different components, or else we
continue with the component of $G'$ that contains all the terminals. 
Suppose there is a guessed subpath in $G'$ containing more than one terminal. If the order is not $t_i, t_{i+1},\ldots, t_j$ for some $i< j$, we discard $G'$. 
Otherwise, if necessary, we place $t_i$ and $t_j$ on this subpath such that they are at distance at least four of each other and also are of distance at least four of each end vertex of the subpath. Because the guessed subpaths are sufficiently long, such a placement is possible. We then remove $t_{i+1},
\ldots, t_{j-1}$ from the list of terminals.
After processing all created graphs as above, we obtain the desired set 
${\mathcal G}$. Since $k$ is fixed, ${\mathcal G}$ can be constructed in polynomial time. 
\end{proof}

\begin{lemma}\label{l-c5}
Let $G$ be a simple claw-free graph.
Removing a vertex $u\in V(G)$, the neighborhood of which contains an induced 
odd hole of length five, preserves the solution. 
\end{lemma}

\begin{proof}
Because $G$ is simple, $u$ is not a terminal.
We first show the following claim.

\medskip
\noindent
{\it Claim 1.}
Let $G[\{v,w,x,y\}]$ be a diamond in which $vw$ is a non-edge. If there is
a solution $P$ that contains $v,x,w$, then there is another solution that
contains $v,y,w$ (and that does not contain $x$). 

\medskip
\noindent
In order to see this take the original solution $P$ and
notice that by claw-freeness any neighbor of $y$ on $P$ must be in the
(closed) neighborhood of $v$ or $w$. This way the solution can be rerouted
via $y$, without using $x$. This proves Claim 1.

\medskip
\noindent
Now suppose that $u$ is a vertex which has an odd hole $C$ of length five in its
neighborhood. Obviously, $G$ is a {\sc Yes}-instance if $G-u$ is a 
{\sc Yes}-instance. 
To prove the reverse implication, suppose $G$ is a {\sc Yes}-instance.
Let $P$ be a solution. If $u$ does not belong to $P$ then we are done.
Hence, we suppose that $u$ belongs to $P$ and consider
three cases depending on $|V(C)\cap V(P)|$.

\noindent
{\it Case 1.} $|V(C)\cap V(P)|\geq 2$.
Then $|V(C)\cap V(P)|=2$, as any vertex on $P$ will have at most two neighbors.
We are done by Claim 1.

\noindent
{\it Case 2.} $|V(C)\cap V(P)|=1$.
Let $w\in V(C)$ belong to $P$ and 
let the other neighbor of $u$ that belongs to $P$ be $x$. 
We note that $x$ must be adjacent to at least one of
the neighbors of $w$ in $C$. Then we can apply Claim 1 again.

\noindent
{\it Case 3.} $|V(C)\cap V(P)|=0$.
Let the two neighbors of $u$ on $P$ be $x$ and $y$. To avoid a claw at $u$,
every vertex of $C$ must be adjacent to $x$ or $y$. If there is a
vertex in $C$ adjacent to both, we apply Claim 1. Suppose
there is no such vertex and that the vertices of the $C$ are
partitioned in two sets $X$ (vertices of $C$ only adjacent to $x$) and $Y$ (vertices of $C$ only adjacent to $y$). 
We assume without loss of generality
that $|X|=3$, and hence contains a pair of independent
vertices which together with $u$ and $y$ form a claw. This is a contradiction.
\end{proof}

\begin{lemma}\label{l-hc}
Let $G$ be a simple quasi-line graph with a homogeneous clique $A$.
Then contracting $A$ to a single vertex preserves the solution and the resulting graph is a simple quasi-line graph containing the same terminals as $G$.
\end{lemma}

\begin{proof}
Each vertex in $A$ lies on a triangle, unless $G$ is isomorphic to $P_2$, 
which is not possible.
Hence, by condition (i), $A$ does not contain a terminal.
We remove all vertices of $A$ except one. The resulting graph will be a
simple quasi-line graph containing the same terminals, and we will preserve the solution.
\end{proof}

\begin{lemma}\label{l-noterminal}
Let $G$ be a simple quasi-line graph with terminals ordered $t_1,\ldots,t_k$
that has no homogeneous clique.
Contracting the cliques $A$ and $B$
in a homogeneous pair 
to single vertices preserves the solution. The resulting graph is quasi-line; 
it is simple unless $A$ or $B$ consists of two vertices $u,u'$ with 
$u\in \{t_1,t_k\}$ and 
$d(u',t_i)\leq 3$ for some $t_i\neq u$.
\end{lemma}

\begin{proof}
Because $G$ does not contain a homogeneous clique,
$V(A)$ and $V(B)$ must be adjacent.
Then, due to condition (ii), there can be at most one terminal in 
$V(A)\cup V(B)$.
In all the cases discussed below we will actually not contract edges but only remove vertices from $A$ and $B$.  
Hence, the resulting graph will always be a quasi-line graph. 

Suppose $A$ contains $t_1$ or $t_k$, say $t_1$. 
Suppose $|V(A)|=1$, so $A$ only contains $t_1$. 
Then the neighbor of $t_1$ is in $B$ and $|V(B)|\geq 2$.  
We delete all vertices from $B$ except this neighbor, because they will
not be used in any solution. Clearly, the resulting graph is simple and the solution is preserved.
Suppose $|V(A)|\geq 2$. Because $t_1$ is of degree one, $A$ consists of
two vertices, namely $t_1$ and its neighbor $t_1'$. 
Note that $t_1'$ does not have a neighbor outside $A$ and $B$, as $t_1$ is of
degree one.
As $V(A)$ and $V(B)$ are adjacent,
$t_1'$ has a neighbor $u$ in $B$. 
We delete $t_1$ and
replace it by $t_1'$ in the set of terminals.
We delete all vertices of $B$
except $u$, because of the following reasons.
If these vertices are not adjacent to $t_1'$, they will never appear 
in any solution. If they are adjacent to $t_1'$, they will not appear
in any solution together with $u$, and as such they can be replaced by $u$. 
Note that $t_1'$ has degree one in the new graph and that this graph is only
simple if $d(t_1',t_j)\geq 4$ for all $2\leq j\leq k$.
Clearly, the solution is preserved.

Suppose $A$ contains a terminal $t_i$ for some $2\leq i\leq k-1$.
Suppose $A$ only contains $t_i$. Because $V(A)$ and $V(B)$ are adjacent,
$t_i$ is adjacent to a vertex $u$ in $B$. 
By condition (i), $u$ is the only vertex
in $B$ adjacent to $t_i$. We delete all vertices of $B$ except $u$.
Clearly, the resulting graph is simple and the solution is preserved.
Suppose $|V(A)|\geq 2$. By condition (ii), $A$ contains only one
other vertex $t_i'$ and $t_i,t_i'$ do not have a common neighbor. 
Then $A$ must be separated of the rest of the graph by $B$. Furthermore, 
the other neighbor of $t_i$ must be in $B$.
We delete $t_i'$ and all vertices of $B$ except the neighbor of $t_i$.
Clearly, the resulting graph is simple and the solution is preserved.

Suppose $A$ does not contain a terminal. By symmetry, we may assume that 
$B$ does not contain a terminal either.
Let $a'b'\in E(G)$ with $a'\in V(A)$ and $b'\in V(B)$. 
Let $G'$ be the graph obtained from 
$G$ by removing all vertices of $A$ except $a'$ and $B$ except $a',b'$. 
Note that we have kept all terminals and that
the resulting graph is simple.
Any solution $P'$ for $G'$ is a solution for $G$.

Now assume we have a solution $P$ for $G$. 
We claim that $|P\cap A|\leq 1$ and $|P\cap B|\leq 1$.
Suppose otherwise, say $|P\cap A|\geq 2$. Then $|P\cap A|=2$, as $P$ is a path.
Since $t_1$ and $t_k$ are not in $A$, we find that $P$ contains a
subpath $xuvy$ with $u,v\in A$.
Since $x$ is adjacent to $u\in A$, but also non-adjacent to $v\in A$, we find that $x\in B$. 
Analogously we get that $y\in B$. However, then $xy\in E(G)$. This is a contradiction.

Suppose $|P\cap A|=0$ and $|P\cap B|=0$. 
Then $P$ is a solution for $G'$ as well. 
Suppose $|P\cap A|=0$ and $|P\cap B|=1$.
Then we may without loss of generality assume
that $b'\in V(P)$ and find that $P$ is a solution for $G'$ as well. 
The case $|P\cap A|=1$ and $|P\cap B|=0$ follows by symmetry.
Suppose $|P\cap A|=|P\cap B|=1$, 
say $P$ intersects $A$ in $a$ and $B$ in $b$.
If $ab \in E(G)$ then we replace $ab$ by $a'b'$ and obtain a solution for $G'$.
Suppose $ab\notin E(G)$. Because $a$ is not a terminal, $a$ has neighbors $x$ and $y$ on $P$.
If $x,y\notin N(b)$ then $\{a',x,y,b'\}$ induces a claw in 
$G$ with center $a'$. This is not possible. 
Hence, we may assume without loss of generality that $y$ is adjacent to $b$.
Since $A$ or $B$ contains at least two vertices, $y$ has degree at least three.
Then $y$ is not a terminal.
Thus we can skip $y$ and exchange $ayb$ in $P$ with $a'b'$ to get the desired 
induced path $P'$.
\end{proof}

\begin{theorem}\label{t-lin}
The {\sc Ordered-$k$-in-a-Path} problem
can be solved in polynomial time in linear interval graphs.
\end{theorem}

\begin{proof}
Let $G$ be a linear interval graph. We may assume without loss of
generality that the terminals form an independent set.
We use its linear representation that we obtain in polynomial time
by Lemma~\ref{l-king2}.
In what follows the notions of predecessors (left) and successors (right) 
are considered for the linear ordering of the points on the line. 
Without loss of generality we may assume that $t_1$ is the first
point and that $t_k$ is the last and that no two points coincide.
By our assumption, $t_i$ and $t_{i+1}$ are nonadjacent.
From the set of points belonging to
the closed interval $[t_i,t_{i+1}]$ we remove all neighbors of $t_i$ except the rightmost one and all
neighbors of $t_{i+1}$ except the leftmost. Then the shortest path between $t_i$ and $t_{i+1}$ is induced.
In addition, these partial paths combined together provide a solution unless
for some terminal $t_i$ its leftmost predecessor is adjacent to its rightmost successor. Hence, no induced path may have $t_i$ among its inner vertices.
\end{proof}

\begin{lemma}[proof postponed to journal version]\label{l-tol}
Let $G$ be a composition of linear interval strips. 
It is possible to create in polynomial time a line
graph $G'$ with $|V(G')|\leq |V(G)|$, while preserving the solution.
\end{lemma}

\begin{theorem}\label{t-linegraph}
For fixed $k$, {\sc Ordered-$k$-in-a-Path} 
is polynomially solvable in line graphs.
\end{theorem}

\begin{proof}
A version of {\sc Ordered-$k$-in-a-Path} in which the path is not necessarily induced can be easily translated into an instance of the {\sc $k$-Disjoint Paths} problem and solved in polynomial time due to Theorem \ref{t-RS}. Noting that mutually induced paths in a line graph $L(G)$  are in one-to-one correspondence with vertex-disjoint paths in $G$ enables us to solve the {\sc Ordered-$k$-in-a-Path} problem in polynomial time for line graphs. 
\end{proof}

\section{Conclusions and Further Research}\label{s-con}
 
We showed that, for any fixed $k$, the problems
{\sc $k$-in-a-Path}, {\sc $k$-Disjoint Induced Paths} and {\sc $k$-Induced Cycle} are polynomially solvable on claw-free graphs.
If $k$ is part of the input these problems are known to be \NP-complete. 
In the journal version 
we show this is true, even when the input is restricted to be claw-free.
Perhaps the two most fascinating related open problems are to determine the complexity of deciding if a graph contains an odd hole (whereas the problem of finding an even hole is polynomially solvable~\cite{CKS05}) and to determine the computational complexity of deciding if a graph contains two mutually induced holes (whereas it is known that the case of two mutually induced odd holes 
is \NP-complete~\cite{GKPT09}). For claw-free graphs 
these two problems are solved.
Shrem et al.~\cite{shrem} even obtained a polynomial-time algorithm for detecting a shortest odd hole in a claw-free graph. 
In the journal version we will address the second problem for claw-free graphs.


\begin{thebibliography}{10}

\bibitem{Bi91}
D.~Bienstock.
On the complexity of testing for odd holes and induced odd paths.
{\em Discrete Mathematics} {\bf 90} (1991) 85--92, 
See also Corrigendum, Discrete Mathematics {\bf 102} (1992) 109.

\bibitem{CKS05}
M.~Chudnovsky, K.~Kawarabayashi and P.D.~Seymour.
Detecting even holes.
{\em Journal of Graph Theory} {\bf 48} (2005) 85--111.

\bibitem{CS05}
M. Chudnovsky and P.D. Seymour.
The structure of claw-free graphs. 
In {\it Surveys in combinatorics 2005}, 
Cambridge (2005) 153--171.

\bibitem{CS}
M.~Chudnovsky and P.D.~Seymour. 
The three-in-a-tree problem. 
{\em Combinatorica}, to appear.
 
\bibitem{DHH96}
X. Deng, P. Hell, and J. Huang.
Linear time representation algorithm for proper circular-arc graphs and proper interval graphs.
{\em SIAM Journal on Computing} {\bf 25} (1996) 390--403.

\bibitem{derhy2009}
N.~Derhy and C.~Picouleau.
Finding induced trees.
{\em Discrete Applied Mathematics} {\bf 157} (2009) 3552--3557. 

\bibitem{DPT09}
N. Dehry, C. Picouleau, and N. Trotignon. The four-in-a-tree problem 
in triangle-free graphs. 
{\em Graphs and Combinatorics} {\bf 25} (2009) 489--502. 

\bibitem{FFR97}
R. Faudree, E. Flandrin, and Z. Ryj{\'a}{\v{c}}ek. 
Claw-free graphs---a survey.
{\em Discrete Mathematics} {\bf 164} (1997) 87--147.

\bibitem{Fe89}
M.R. Fellows.
The RobertsonÐSeymour theorems: A survey of applications.
In:
{\it Proceedings of AMS-IMS-SIAM Joint Summer Research 
Conf. Contemporary Mathematics}, 
Providence, RI (1989) 1-Ð18.

\bibitem{GKPT09}
P. Golovach, M. Kami\'nski, D. Paulusma, and D. M. Thilikos.
Induced packing of odd cycles in a planar graph.
In: {\em Proceedings of ISAAC 2009}, LNCS 5878 (2009) 514--523.

\bibitem{HH06}
R. Haas and M. Hoffmann.
Chordless paths through three vertices.
{\em Theoretical Computer Science} {\bf 351} (2006) 360--371.

\bibitem{HKP09}
P. van 't Hof, M. Kami\'nski and D. Paulusma.
Finding induced paths of given parity in~claw-free graphs.
In: {\em Proceedings of WG 2009}, LNCS, to appear.


\bibitem{KR08}
A. King and B. Reed. Bounding $\chi$ in terms of $\omega$ and $\delta$ 
for quasi-line graphs.
{\em Journal of Graph Theory} {\bf 59} (2008) 215-Ð228. 

\bibitem{KK08}
Y. Kobayashi and K. Kawarabayashi.
The induced disjoint paths problem.
In: {\em Proceedings of IPCO 2008}, LNCS 5035 (2008) 47--61.

\bibitem{KK09}
Y. Kobayashi and K. Kawarabayashi.
Algorithms for finding an induced cycle in planar graphs 
and bounded genus graphs.
In: {\em Proceedings of SODA 2009} (2009) 1146--1155.

\bibitem{LMT07}
B.~L\'{e}v\^{e}que, D.Y.~Lin, F.~Maffray, and N.~Trotignon.
Detecting induced subgraphs.
{\em Discrete Applied Mathematics} {\bf 157} (2009) 3540--3551.

\bibitem{RS95}
N.~Robertson and P.D.~Seymour.
\newblock Graph minors. XIII. The disjoint paths problem.
\newblock {\em Journal of Combinatorial Theory, Series B} {\bf 63} (1995) 65--110.

\bibitem{CRST06}
M.~Chudnovsky, N.~Robertson, P.D.~Seymour, and R.~Thomas.
The strong perfect graph theorem.
{\it Annals of Mathematics} {\bf 164} (2006) 51--229.

\bibitem{shrem}
S.~Shrem, M.~Stern and M.C.~Golumbic.
Smallest odd holes in claw-free graphs.
In {\em Proceedings of WG 2009}, LNCS 5911 (2009) 329--340. 

\bibitem{TW09}
N. Trotignon and L. Wei.
The $k$-in-a-tree problem for graphs of girth at least $k$,
manuscript.

\end{thebibliography}
\end{document}